# Quantum mechanics as a result of time broadening of the classical object

A. M. Ghorbanzadeh

*Department of physics, Sharif University of Technology, P.O.Box:11365-9161
Tehran, Iran,, E-mail: ata1@sharif.edu*

It is shown that the Schrödinger equation is a byproduct of more deterministic Boltzmann-like equation. All physical information is derived from the solution of this equation, which is a function of space and momentum. The additional terms in this equation, when compared with the classical Boltzmann equation, asserts that the particle should be broadened over the time axis, i.e., it is partly present at the past, the present, and the future. The solution of this equation itself is nonlocal. This solution requires an especial time coordinate that is different from the usual time parameter entering into the formalism of quantum mechanics. This time coordinate is hidden in the structure of wave function. It is substantiated that the mixed state is a natural result of time broadening. Furthermore, the interaction between an object that is broadened in time and an observer that has a very narrow time-width leads to the collapse of wave function.

## I.   INTRODUCTION

Without any doubt the quantum mechanics has had one of the greatest impacts on the science and technology amongst the human discoveries. Nevertheless, its great success has been accompanied by our inability of giving a proper interpretation to it. However, after about one century of inadequate or imperfect alternative theories, the majority numbers of scientists have used it safely to apply to many nowadays problems and set aside the philosophy behind it. But, it is still the cause of discomfort for those who think solving this problem may radically change our understanding of the nature. The origin of this dissatisfaction, as Einstein thought, is obviously due to the indeterminism at the heart of quantum mechanics. The instructions contained in the Copenhagen interpretation even prohibits one to move further, as if, it is the last station of human speculation. The most highlights of different efforts have been made to clarify the issue, during last decades, are the Copenhagen interpretation led by Bohr, the proposal of EPR experiments [1], Bell's inequality and its experimental test [2,3], Schrödinger's cat paradox [4], Von Neumann's impossibility proof and his conscious observer theory [5], quantum potential of David Bohm [6], the many worlds scenario of Everett [7], GRW of Ghirardi [8], and transactional quantum theory [9]. Despite these efforts, it is difficult to say that the problem already has been solved. The desperation of lacking an adequate theory was so pronounced that even the intervention of hand of the God has been proposed [10]. Nevertheless, if all these theories have not yet fully provided the answer, they have at least addressed the conceptual problems very well.

In this paper a model that is more general than the usual quantum mechanics is introduced. It is based on a new insight into the concept of time. The time is a parameter which determines the evolution of any system. More precisely, assuming a system is located at some point $t_0$ on time axis, after a while it will evolve to a new point $t$. We do not usually ask how sharp this location is on the time axis. But, this question will appear to be important for reasons described below. Let's try to imagine a creature that partly feels itself present at all points on the time axis, i.e. the past, the present, and the future. What would its evolution mean? Since all things in interaction with it seem stationary, it would be evolution free. Meanwhile, the other aspect of this situation would sound more interesting. This phenomenon is depicted in Fig.1. The observer sees three distinct events in different times of the past, the present, and the future successively in frames 1(a)-1(c) of Fig.1. Now suppose, for some reason, the observer is broadened in time as he would cover many points on the time axis. Therefore, the picture 1d would be what he will observe. The same picture would be resulted if the thing to be observed could be broadened in time and the usual observer could equally interact with it. At first glance, an unusual state is evident in the picture 1d. It promptly reminds the mixed state in quantum mechanics. Is this resemblance accidental? The answer to this question, which is the subject of this paper, will be negative. These two pictures are quite the same, i.e., the quantum mechanics and the mechanics resulted from the time broadened objects.

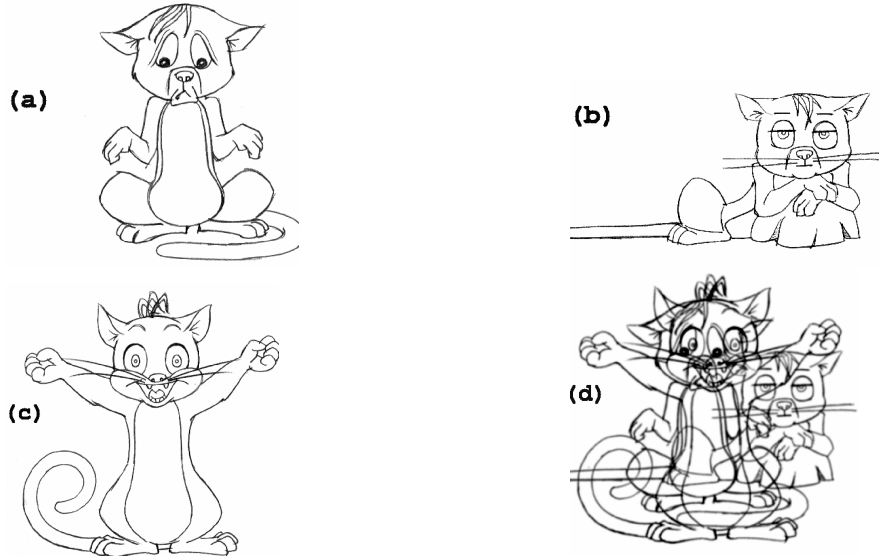

**Fig. 1** Three distinct events at the past (a), the present (b), the future (c), and the broadened time state (d)

## II. THE EQUIVALLENCE OF SHRODINGER AND BOLTZMANN EQUATIONS

The dynamic of non-relativistic particles is explained by the well-known Schrödinger equation:

$$(-\frac{\hbar^2}{2m}\nabla^2 + V(r))\psi = i\hbar \frac{\partial}{\partial t}\psi \qquad (1)$$

The wave function $\psi$ in conjunction with some other operators represents all physical quantities which could be known by performing measurement. To better understand the physical meaning of this function, it is decomposed into two real quantities:

$$\psi = \rho^{1/2} \exp(i\phi) \qquad (2)$$

By inserting Equ. (2) in Equ.(1), two relations are obtained:

$$\partial_t \rho + \nabla \cdot [\rho(\frac{\hbar \nabla \phi}{m})] = 0 \qquad (3)$$

$$-\hbar \partial_t \phi = V(r) + \frac{m}{2}(\frac{\hbar \nabla \phi}{m}) \cdot (\frac{\hbar \nabla \phi}{m}) + \varepsilon \qquad (4)$$

where, $\varepsilon$ is a function of $\rho$ only

$$\varepsilon = \frac{-\hbar^2}{2m\rho}(\nabla^2 \rho - \frac{(\nabla \rho)^2}{2\rho}) \qquad (5)$$

We followed the procedure that was introduced first by Bohm [6]. The first relation shows the continuity equation. All terms in the second equation have dimension of energy, hence it may represent the definition of the system's total energy. At this point, as will be substantiated in what follows, in contrary to the Bohm's version who stated $\varepsilon$ as a potential, we will have another interpretation for it. By the help of the relations (3) and (4), and after some algebra, the following relation is also derived:

$$\partial_t [m\rho(\frac{\hbar \nabla \phi}{m})] + \nabla \cdot [m\rho(\frac{\hbar \nabla \phi}{m})(\frac{\hbar \nabla \phi}{m})] - \rho F + \rho \nabla \varepsilon = 0 \qquad (6)$$

where $F = -\nabla V(r)$. Equations (3) and (6) recall the so called "moments" of the well-known Boltzmann equation for a classical system (see Appendix A). We could continue this process and obtain infinite number of equations that all have the same information that the Schrödinger equation has. The resemblance of Eqs. (3) and (6) with relations (A1) and (A2), respectively, and a great similarity of Eqs. (4) and (A8) suggest the following equivalent relations of quantum parameters with their classical counterparts (see Appendix A)

$$\mathbf{u} \equiv \hbar \nabla \phi / m \qquad (7)$$

$$e + V(\mathbf{r}) = \varepsilon_c + \frac{1}{2}mu^2 + V(\mathbf{r}) \equiv -\hbar \partial_t \phi, \Rightarrow \varepsilon \equiv \varepsilon_c \qquad (8)$$

and also

$$\nabla \cdot \mathbf{P} \equiv \rho \nabla \varepsilon \qquad (9)$$

might be satisfied. As will be shown later, the equivalence relation (9) is an approximation. Therefore, the consistent interpretation of $\varepsilon$ appears to be the average internal kinetic energy.

Hence, the quantum mechanics behave very much like the classical kinetics, and so, can be similarly deduced from some unknown more deterministic "quantum distribution function $f(\mathbf{r}, \mathbf{p}, t)$". It asserts that a quantum particle has different velocities at a given point, the occurrence (probability) of which is given by this distribution function.

What is indeed this distribution function? The starting point will be from the prediction of quantum mechanics for the probability of occurrence of the momentum $p$ which is determined by the product of two kets $|\psi\rangle$ and $|\mathbf{p}\rangle$:

$$f(\mathbf{p}) = \langle \mathbf{p}|\psi\rangle\langle\psi|\mathbf{p}\rangle = [1/(2\pi\hbar)^3]\int \psi(\mathbf{r},t)\psi^*(\mathbf{r}',t)\exp[-\frac{i}{\hbar}\mathbf{p}.(\mathbf{r}-\mathbf{r}')]d^3\mathbf{r}d^3\mathbf{r}'$$

or

$$f(\mathbf{p}) = \int \text{Re}\, Q(\mathbf{r},\mathbf{p},t)d^3\mathbf{r} \qquad (10)$$

The $Q(\mathbf{r},\mathbf{p},t)$ is a complex function depending on the wave function $\psi(\mathbf{r},t)$ and $\mathbf{p}$

$$Q(\mathbf{r},\mathbf{p},t) = [1/(2\pi\hbar)^3]\psi(\mathbf{r},t)\int \psi^*(\mathbf{r}',t)\exp[-\frac{i}{\hbar}\mathbf{p}.(\mathbf{r}-\mathbf{r}')]d^3\mathbf{r}'$$

$$= \frac{1}{(2\pi\hbar)^{3/2}}\psi(\mathbf{r},t)\exp(-\frac{i}{\hbar}\mathbf{p}.\mathbf{r})\phi^*(\mathbf{p}) \qquad (11)$$

Where $\phi(\mathbf{p})$ is the Fourier transform of $\psi(\mathbf{r},t)$. In Eq. (10) $f(\mathbf{p})$ is real and the application of the operator Re does not change any.

Conversely, by the definition of the quantum distribution function $f(\mathbf{r},\mathbf{p},t)$ it is expected that:

$$f(\mathbf{p}) = \int f(\mathbf{r},\mathbf{p},t)d^3\mathbf{r} \qquad (12)$$

Note that everywhere appropriate $\mathbf{p}$ or $\mathbf{v} = \mathbf{p}/m$ is used as the independent variable. Comparison of Eqs. (10) and (12) yields:

$$f(\mathbf{r},\mathbf{p},t) = \text{Re}[Q(\mathbf{r},\mathbf{p},t)] \qquad (13)$$

This function is normalized:

$$\int \text{Re}[Q(\mathbf{r},\mathbf{p},t)]d^3\mathbf{p}d^3\mathbf{r} = \int |\psi|^2 d^3\mathbf{r} = 1 \qquad (14)$$

By knowing the distribution function, the average velocity of the quantum system is obtained

$$\mathbf{u} = \langle \mathbf{p}/m \rangle = \int f(\mathbf{r},\mathbf{p},t)(\mathbf{p}/m)d^3\mathbf{p} \Big/ \int f(\mathbf{r},\mathbf{p},t)d^3\mathbf{p}$$

$$= (1/m\rho)\text{Re}[\psi(\mathbf{r},t)(i\hbar)\nabla\psi^*(\mathbf{r},t)] = \hbar\nabla\varphi/m \qquad (15)$$

The total kinetic energy is

$$k = \langle \mathbf{p}^2/2m \rangle = \int \text{Re}[Q(\mathbf{r},\mathbf{p},t)](\mathbf{p}^2/2m)d^3\mathbf{p} \Big/ \int \text{Re}[Q(\mathbf{r},\mathbf{p},t)]d^3\mathbf{p}$$

$$= (1/2m\rho)\text{Re}[(i\hbar)^2\psi\nabla^2\psi^*] = \frac{-\hbar^2}{2m\rho}(\nabla^2\rho - (\nabla\rho)^2/2\rho) + (\hbar\nabla\varphi)^2/2m \qquad (16)$$

The relations (15) and (16) are just the regenerations of Eqs. (7) and (4), or (5), that were independently suggested from the Schrödinger equation. It seems that the consistent distribution function has been explored.

Before proceeding further, two important features of the quantum distribution function are discussed. The equation (13) can be re-expressed as:

$$f(\mathbf{r},\mathbf{p},t) = \text{Re}[Q(\mathbf{r},\mathbf{p},t)] = \text{Re}[q(\mathbf{r},\mathbf{p},t)\int q^*(\mathbf{r}',\mathbf{p},t)d^3\mathbf{r}'] \qquad (17)$$

where

$$q(\mathbf{r},\mathbf{p},t) = [1/(2\pi\hbar)^{3/2}]\psi(\mathbf{r},t)\exp(-(i/\hbar)\mathbf{p}.\mathbf{r}) \qquad (18)$$

First, the value of the distribution function at time $t$ and at location $\mathbf{r}$ is the sum of all contributions that come from all other points $\mathbf{r}'$ through function $q^*(\mathbf{r}',\mathbf{p},t)$, which represents some local property of the matter. Furthermore, from the definition (17) it is apparent that this function is nonlocal, in that every change that is made in it at the time $t$ and at distant location $\mathbf{r}'$ is promptly reflected in the distribution function at $\mathbf{r}$, again, through $q^*(\mathbf{r}',\mathbf{p},t)$. Otherwise, $q^*(\mathbf{r}',\mathbf{p},t)$, in the expression (17), should have looked as $q^*(\mathbf{r}',\mathbf{p},t-|\mathbf{r}-\mathbf{r}'|/c')$ where $c'$ would be the velocity of perturbation propagation. It behaves as if the part of particle existing at $\mathbf{r}'$ acts as an independent source and propagates with infinite velocity to other points. It is noted that the nonlocality is a major necessity for every theory to interpret the measurement.

By direct differentiation, it can be shown that $Q(\mathbf{r},\mathbf{p},t)$ satisfies the following relation:

$$\partial_t Q + \frac{1}{m}\mathbf{p}.\nabla Q$$
$$= \frac{i\hbar}{2m}\nabla^2 Q + \frac{1}{i\hbar}[V(\mathbf{r}) - \frac{1}{(2\pi\hbar)^3}\psi(\mathbf{r},t)\int \psi^*(\mathbf{r}',t)V(\mathbf{r}')\exp(-i\frac{\mathbf{p}}{\hbar}.(\mathbf{r}-\mathbf{r}')d^3\mathbf{r}'] \qquad (19)$$

The definition of $Q(\mathbf{r},\mathbf{p},t)$, see Eq. (11), helps to rewrite the last term of above equation as:

$$\frac{Q(\mathbf{r},\mathbf{p},t)}{|\varphi(\mathbf{p})|^2}\int V(\mathbf{r}')Q^*(\mathbf{r}',\mathbf{p},t)d^3\mathbf{r}' = \langle V(\mathbf{r})\rangle Q(\mathbf{r},\mathbf{p},t) \qquad (20)$$

The statistical average $\langle V(\mathbf{r})\rangle$ is determined as below:

$$\langle V(\mathbf{r})\rangle = \frac{1}{\int Q(\mathbf{r},\mathbf{p},t)d^3\mathbf{r}}\int V(\mathbf{r}')Q^*(\mathbf{r}',\mathbf{p},t)d^3\mathbf{r}'$$

reminding that:

$$\int Q(\mathbf{r},\mathbf{p},t)d^3\mathbf{r} = |\varphi(\mathbf{p})|^2$$

Therefore, the complex quantum distribution function $Q(\mathbf{r},\mathbf{p},t)$ is the answer of Hamiltonian-like equation:

$$H_b Q(\mathbf{r},\mathbf{p},t) = i\hbar \frac{D}{Dt}Q(\mathbf{r},\mathbf{p},t) \qquad (21)$$

In the above equality, $H_b$ and fluid derivative $\frac{D}{Dt}$ are defined respectively by the following operators:

$$H_b = -\frac{\hbar^2}{2m}\nabla^2 + V(\mathbf{r})_{eff} \ , \ \frac{D}{Dt} = \frac{\partial}{\partial t} + \frac{\mathbf{p}}{m}.\nabla$$

where the effective potential is:

$$V(\mathbf{r})_{eff} = V(\mathbf{r}) - \langle V(\mathbf{r})\rangle$$

Again, the moving particle nonlocally sees the potential of all points, as if it spread all over the space. The Eq. (19) takes more familiar form by using the Taylor expansion of $V(\mathbf{r})$:

$$V(\mathbf{r}') = V(\mathbf{r}) + \sum_1^\infty \frac{1}{n!}\nabla^n V(\mathbf{r}):(\mathbf{r}'-\mathbf{r})^n$$

or

$$\langle V(\mathbf{r})\rangle Q = [V(\mathbf{r}) + \sum_1^\infty \frac{(-i\hbar)^n}{n!}\nabla_\mathbf{r}^n V(\mathbf{r}):\nabla_\mathbf{p}^n]Q$$

where the equality

$$(\mathbf{r}' - \mathbf{r})^n = (-i\hbar)^n \nabla_{\mathbf{p}}^n$$

was used.
By inserting it into the Eq. (19), a more familiar relation is obtained

$$\partial_t Q + \frac{1}{m}\mathbf{p}.\nabla Q + \mathbf{F}.\nabla_{\mathbf{p}} Q = \frac{i\hbar}{2m}\nabla^2 Q + \sum_{n=2}^{\infty} \frac{(-i\hbar)^{n-1}}{n!}\nabla_{\mathbf{r}}^n V(\mathbf{r}) : \nabla_{\mathbf{p}}^n Q \qquad (22)$$

By taking the Re both sides of the Eq. (22), a Boltzmann-like equation, which hereinafter we call it the quantum Boltzmann equation, is obtained for quantum distribution function $f(\mathbf{r},\mathbf{p},t)$:

$$\partial_t f + \frac{1}{m}\mathbf{p}.\nabla f + \mathbf{F}.\nabla_{\mathbf{p}} f = -\frac{\hbar}{2m}\nabla^2 \,\text{Im}\, Q(\mathbf{r},\mathbf{p},t)$$

$$+ \sum_{n=3(odd)}^{\infty} (-1)^{(n-1)/2}(\hbar)^{n-1}\frac{1}{n!}\nabla_{\mathbf{r}}^n V(\mathbf{r}) : \nabla_{\mathbf{p}}^n f$$

$$+ \sum_{n=2(even)}^{\infty} (-1)^{(n-2)/2}(\hbar)^{n-1}\frac{1}{n!}\nabla_{\mathbf{r}}^n V(\mathbf{r}) : \nabla_{\mathbf{p}}^n \,\text{Im}\, Q \qquad (23)$$

At the classical limit, $\hbar \to 0$, Eq. (23) is converted to the well-known Boltzmann equation for one particle:

$$\partial_t f + \frac{1}{m}\mathbf{p}.\nabla f + \mathbf{F}.\nabla_{\mathbf{p}} f = 0 \qquad (24)$$

In analogy with the classical Boltzmann equation, this equation should give all conservation relations including those suggested directly from Schrödinger equation, say, conservation of mass, momentum, and the energy. As it has been shown in the Appendix B, it exactly gives the same conservation relations. Therefore, this kinetic model seems to operate well and gives all predictions of quantum mechanics in a consistent way. Conversely, we could start from it and, as a by product, obtain the Schrödinger equation and conventional quantum mechanics.

The second and the third terms at the right hand side of Eq. (23) are similar to the Fokker-Planck representation of weak collisions. The interpretation of collision for these terms might have been true for many particle systems. But, if the system contains only one particle, as it is here, interpreting them as collision would require the interaction of particle with itself or something else that should be accepted with cares. The difficulty arising in understanding the additional terms in the quantum Boltzmann equation when it is compared with the classical Boltzmann relation for one particle (24), may be understood better when we trace the Boltzmann equation back to the Liouville equations, from which it is principally derived. The Liouville equation assumes that the number of systems, regardless of which volume they occupy in the phase space during time evolution, is constant. This assumption leads to zero at the right hand side of the Boltzmann equation (24) for one particle (see Appendix C). The appearance of new terms in Eq. (23), forces us to reconsider this, seemingly, trivial assumption. The quantum mechanics attacks this assumption. There is no way remained except for assuming that a quantum system behaves like a chalk on the blackboard, retaining some of its part in the past during time evolution towards the future in the phase space, provided that we insist on what we understand of the velocity and force as those introduced in (C3). Broadening over the time axis is equivalent to the violation of number of systems conservation during the evolution. It also makes a non-vanishing amount at the right hand side of the Liouville equation. This is indeed consistent with the assumption of nonlocality. Thus, for a free particle the term

$$(-\hbar/2m)\nabla^2 \text{Im} Q(\mathbf{r},\mathbf{p},t)$$

is a measure of "time broadening". It is roughly estimated as,

$$\frac{\hbar f}{2md^2} \approx \frac{\hbar(1/\Delta V)}{2md^2} \approx \frac{\hbar}{2md^5}$$

where $\Delta V \approx d^3$ is the dimension over which the presence of the particle is probable. For an electron confined in a potential barrier of a nanometer size, this amounts to a value of $10^{41} s^{-1} m^{-3}$. While, for a $cm$ size object with mass of gram it is about $10^{-21} s^{-1} m^{-3}$. Therefore, the conventional objects are quite localized in time.

Furthermore, $\text{Im} Q(\mathbf{r},\mathbf{p},t)$ that is responsible for time broadening, also satisfies an equation which is very similar to the particle equation (23) and, thus, may represent a coexisting particle. But, it's integral over the space or momentum is zero, so it cannot objectively exist.

Now, again we come back to refine our understanding of the quantum distribution function that, by Eq. (17), causes the instantaneous propagation of perturbations. It was, classically, expected that every local change in the distribution function would have propagated by a speed, specified by the particle's local velocity, to the distant places, namely, the future. Without the necessity for the assumption of infinite velocity, this instantaneous effect can be attributed to the time broadening of the particle. The distance between two points $a$ and $b$ is traveled by a classical particle during a finite time interval $\Delta t$. Conversely, the particle that is extended over the past, the present, and the future, not needing to care the classical rules, promptly feels a part of its existence at the future, i.e., the point $b$. This meaning is distinguished from the particle disintegration. By this new insight, the problems of quantum mechanics will be treated.

## III. THE MIXED STATE

All problems in quantum mechanics arise from, the classically unprecedented, mixed state. To understand this unusual state of matter the example of pair particle production is considered. Suppose two identical particles such as electrons are generated. They can be produced by double ionization of atoms or by other means. Except for some constraints that should be observed, that are the conservation of total momentum and net angular momentum that both are supposed to be zero, we have no control over them. The direction of flight, at the center of mass frame, could be arbitrary. Suppose at the initial time the electrons are ejected at a specified angle. Since the electrons are broadened in time, at later times they can be partly present at the past, saying, at the generation instance. This leads to a non-stop generation procedure each time generating a new random direction of flight. The resultant will be an isotropic cloud, or the well-known spherical wave. These continuous generations also lead the system to contain all possible spin directions for two particles, for which the singlet state is a representation. Although, one fact which should not be ignored is that despite representing this situation by a single wave function at each time, each direction of ejection or the direction of spin belongs to the distinct "generation times". This generation times are hidden in the wave function $\psi(\mathbf{r},t)$. The parameter $t$ appearing in the quantum mechanics may represent a parameter of evolution and is not the real "time coordinate". Hence, we should be able to express the quantum distribution function as a superposition of more deterministic distribution function $q(\mathbf{r},\mathbf{p},t,t_h)$ characterized by the hidden time $t_h$. The following definition can be a possibility

$$Q(\mathbf{r},\mathbf{p},t) = \int_{t_0}^{t} q(\mathbf{r},\mathbf{p},t,t_h) dt_h \qquad (25)$$

## IV. THE MEASUREMENT

The local function $q(\mathbf{r},\mathbf{p},t)$, introduced in Eq. (18), verifies the next Hamiltonian-like equation:

$$i\hbar(\frac{\partial}{\partial t} + \frac{\mathbf{p}}{m}.\nabla)q(\mathbf{r},\mathbf{p},t) = [-\frac{\hbar^2}{2m}\nabla^2 + V(\mathbf{r}) + \frac{p^2}{2m}]q(\mathbf{r},\mathbf{p},t) \qquad (26)$$

or

$$i\hbar\frac{\partial}{\partial t}q(\mathbf{r},\mathbf{p},t) = [(1/2m)(\frac{\hbar}{i}\nabla + \mathbf{p})^2 + V(\mathbf{r})]q(\mathbf{r},\mathbf{p},t) \qquad (26')$$

It is evident that $q^*(\mathbf{r},\mathbf{p},t)$ is a time-reversed solution of the above equations. Then, by referring to the integral definition (17), the quantum distribution function $Q(\mathbf{r},\mathbf{p},t)$ is specified by the product of a quasi-plane wave $q(\mathbf{r},\mathbf{p},t)$, with momentum p at the location $\mathbf{r}$, and its time reversed at the location $\mathbf{r}'$. Therefore, a possible interpretation is that each point in the space, specified by the domain where the wave function $\psi$ is nonvanishing, can be regarded as a source which symmetrically propagates the matter towards the future and the past. Consequently, the probability of finding the particle at a specified point is determined by two contributions from the time-forwarded $q(\mathbf{r},\mathbf{p},t)$, and time-reversed waves emitted from all points. The equation (17) states that this propagation takes no time. However, if we accept the time broadening, the need of infinite velocity is avoided. So, the mutual relation of two separate points by two time-broadened forward and backward propagations is responsible for the distant nonlocal action. The transactional interpretation of quantum mechanics may, in this regard, have some similarity with this reasoning.

Alternatively, the distribution function can be shown to be

$$f(\mathbf{r},\mathbf{p},t) = \text{Re}[Q(\mathbf{r},\mathbf{p},t)] = \text{Re}\{\int [\rho(\mathbf{r},t)\rho(\mathbf{r}',t)]^{1/2} e^{\frac{i}{\hbar}S(\mathbf{r},\mathbf{r}',\mathbf{p},t)} d^3\mathbf{r}'\} \qquad (27)$$

where

$$S(\mathbf{r},\mathbf{r}',\mathbf{p},t) = \int_{\mathbf{r}'}^{\mathbf{r}} [\mathbf{u}(\mathbf{r}'') - \mathbf{p}].d\mathbf{r}'' \qquad (28)$$

In deriving the above expression, the definitions (2) and (7) for wave function and average velocity $\mathbf{u}$ were used, respectively. The phase factor in (27), containing the action $S$, may give the probability for propagation from $\mathbf{r}'$ to $\mathbf{r}$. It is independent of path.

The measurement is a result of interaction between an object that is broadened in time and an observer that has a very narrow "time-width". To qualitatively explain the measurement mechanism, a specific example of Stern-Gerlagh experiment is examined. In this experiment an electron enters the inhomogeneous magnetic field after which the two spin components are spatially separated. According to the above discussion, these two spatially distinct packets are interrelated by the time-reversed wave propagation. The forward propagation more and more disconnect them. At the instant of observation only one packet has time-coincidence with the time of observer, reminding that the two packets indeed belong to the different hidden times. As a trivial fact we know that we cannot interact with the dinosaurs, because we and dinosaurs belong to the different epochs. The same reasoning says that, the observer with narrow time-width located at the present cannot interact with the past of the time-broadened object being observed. Hence, the observer should see the whole object at the sharp time of the observation. For example, he should measure for the electrical charge of the electron an amount of $e$ not a fraction of it. This is the reason of wave function collapse. The process of observation strongly amplifies the time-reversed propagation of the other packet while suppressing it to be propagated forwardly. The nonlocal nature of the quantum distribution function allows this process to happen instantaneously. Mathematically, the observation corresponds to the inner product of the time kets of the object by the time coordinate ket of the observer, where the amplification process also should be incorporated.

Conversely, if the observer were equally broadened in time no collapse would be resulted. Such observer could see all hidden times. A partial collapse happens when the observer and the object have different time-widths.

At the instant of the observation the distribution function is no longer derived by the definition (13). In fact, by analogy with the classical kinetic theory, the distribution function should pass through a transient nonequilibrium stage, where the system is not explained by a function like $Q(\mathbf{r},\mathbf{p},t)$ or its equivalent $\psi(\mathbf{r},t)$. However, by adding a collision term that is responsible for interaction, the quantum Boltzmann equation (23) should in principle act. In fact, to reach a time broadened solution, such as distribution (13), probably a finite period of time should be passed to permit the build up of the nonlocal solution. This mechanism demands a separate formalism.

## V. CONCLUSION

In summary, a deterministic model was introduced which gives the usual quantum mechanics in a consistent way. The Schrödinger equation is an average over the momentum space of a more general equation. In this regard, the Einstein who believed that the usual quantum mechanics is not the whole story may was right. This model claims that the time is on the same footing as the space coordinates. The object can freely, within a limited extend, change the direction of time and go to the past as well as the future. Hence, the parameter $t$ appearing in the usual quantum mechanics is only a dynamical parameter showing, for example, the maximum available hidden time value, not the time coordinate. The equilibrium solution of the so called quantum Boltzmann equation is nonlocal. The form of the solution requires that any perturbation made to it should be instantaneously propagated over the whole space. We as an observer with the narrow time-width can never taste the whole reality of the object, without perturbing it. This is what Bohr insisted on it. However, before the measurement the reality exists. It is only not accessible to imperfect beings as us. To feel it, we should be, seemingly impossible, broadened in time. But, who knows the future.

## ACKNOWLEDGMENT

The author greatly appreciates M. Seify for his animated pictures of cat.

## APPENDIX A: THE MOMENTS OF THE BOLTZMANN EQUATION

All information of a classical statistical system is provided by a distribution function $f(\mathbf{r},\mathbf{v},t)$, so that $f(\mathbf{r},\mathbf{v},t)d^3\mathbf{r}d^3\mathbf{v}$ shows the probability, for the system, to be in a small neighborhood of the point $(\mathbf{r},\mathbf{v})$. This distribution function satisfies the Boltzmann equation. As known, by multiplying the Boltzmann equation by $\mathbf{v}^n$ and subsequent integration with respect to $\mathbf{v}$, the moments of $f(\mathbf{r},\mathbf{v},t)$ are obtained. Three first moments of $f(\mathbf{r},\mathbf{v},t)$ are the continuity, the conservation of momentum, and the energy conservation (for example see [11]:

$$\partial_t \rho + \nabla \cdot (\rho \mathbf{u}) = 0 \qquad (A1)$$

$$\partial_t (m\rho\mathbf{u}) + \nabla \cdot (m\rho\mathbf{u}\mathbf{u} + \mathbf{P}) - \rho\mathbf{F} = \mathbf{0} \qquad (A2)$$

$$\frac{\partial}{\partial t}(\rho\varepsilon_c + \frac{1}{2}m\rho u^2) + \nabla \cdot [\rho\mathbf{u}(\varepsilon_c + \frac{1}{2}m\rho u^2) + \mathbf{q} + \mathbf{P}\cdot\mathbf{u}] - \frac{\mathbf{F}}{m}\cdot\rho\mathbf{u} = 0 \qquad (A3)$$

In the above relations, $\rho$ is the particle density, $\mathbf{u}$ is the mean velocity, $\varepsilon_c$ is the internal kinetic energy, $\mathbf{P}$ is the tensor of pressure, and $\mathbf{q}$ is the heat flux vector. They are given by:

$$\rho = \int f(\mathbf{r},\mathbf{v},t)d^3\mathbf{v} \qquad (A4)$$

$$\mathbf{u} = \int f(\mathbf{r},\mathbf{v},t)\mathbf{v}d^3\mathbf{v} / \int f(\mathbf{r},\mathbf{v},t)d^3\mathbf{v} \qquad (A5)$$

$$\mathbf{P} = m\int f(\mathbf{r},\mathbf{v},t)(\mathbf{v}-\mathbf{u})(\mathbf{v}-\mathbf{u})d^3\mathbf{v} / \int f(\mathbf{r},\mathbf{v},t)d^3\mathbf{v} \qquad (A6)$$

$$\mathbf{q} = m\int f(\mathbf{r},\mathbf{v},t)(\mathbf{v}-\mathbf{u})(\mathbf{v}-\mathbf{u})^2 d^3\mathbf{v} / \int f(\mathbf{r},\mathbf{v},t)d^3\mathbf{v} \qquad (A7)$$

The total kinetic energy, which contains $\varepsilon_c$ is

$$e = \left\langle \frac{1}{2}mv^2 \right\rangle = \int (\frac{1}{2}mv^2)f(\mathbf{r},\mathbf{v},t)d^3\mathbf{v} / \int f(\mathbf{r},\mathbf{v},t)d^3\mathbf{v}$$

$$= \frac{1}{\rho}\int (\frac{1}{2}m(\mathbf{c}+\mathbf{u})^2)f(\mathbf{r},\mathbf{v},t)d^3\mathbf{v} = \varepsilon_c + \frac{1}{2}m\mathbf{u}^2 \qquad (A8)$$

Where $\mathbf{c} = \mathbf{v}-\mathbf{u}$, and $\varepsilon_c$ is the internal kinetic energy that is related to the temperature. The later is defined by

$$\varepsilon_c = \frac{m}{2}\int f(\mathbf{r},\mathbf{v},t)(\mathbf{v}-\mathbf{u})^2 d^3\mathbf{v} / \int f(\mathbf{r},\mathbf{v},t)d^3\mathbf{v} \qquad (A9)$$

# APPENDIX B: MOMENTS OF THE QUANTUM BOLTZMANN EQUATION

**B1. Conservation of mass**

Before deriving the conservation relations, some preliminary useful mathematical relations are derived

$$\int Q(\mathbf{r},\mathbf{p},t)d^3\mathbf{p} = \rho \tag{B1}$$

Or in general

$$\int Q(\mathbf{r},\mathbf{p},t)\mathbf{p}^n d^3\mathbf{p} = (i\hbar)^n (\psi(\mathbf{r},t)\nabla^n \psi^*(\mathbf{r},t)) \tag{B2}$$

$$\int \mathbf{p}.\nabla f(\mathbf{r},\mathbf{p},t)d^3\mathbf{p} = \text{Re }\nabla.\int \mathbf{p}Q(\mathbf{r},\mathbf{p},t)d^3\mathbf{p} = \text{Re }\nabla.[\psi(\mathbf{r},t)i\hbar\nabla\psi^*(\mathbf{r},t)]$$
$$= \nabla.[\rho(\hbar\nabla\varphi)] \tag{B3}$$

By using (B2) and the definition (2) the next relations are easily derived

$$\int \mathbf{F}.\nabla_p f(\mathbf{r},\mathbf{p},t)\mathbf{p}d^3\mathbf{p} = -\mathbf{F}.\int (\nabla_\mathbf{p}\mathbf{p})f(\mathbf{r},\mathbf{p},t)d^3\mathbf{p} = -\mathbf{F}\rho \tag{B4}$$

The pressure tensor is

$$\mathbf{P} = \text{Re}\int Q(\mathbf{r},\mathbf{p},t)\mathbf{pp}d^3\mathbf{p} = \rho(\hbar\nabla\varphi)(\hbar\nabla\varphi) - \hbar^2 \rho^{1/2}\nabla(\frac{1}{2}\rho^{-1/2}\nabla\rho) \tag{B5}$$

In addition

$$\int \mathbf{F}.\nabla_\mathbf{p} f(\mathbf{r},\mathbf{p},t)d^3\mathbf{p} = \mathbf{F} \cdot f(\mathbf{r},\mathbf{p},t)\big|_{boundary} = 0 \tag{B6}$$

For $n \geq 1$

$$\int \nabla_\mathbf{p}^n Q(\mathbf{r},\mathbf{p},t)d^3\mathbf{p} = \nabla_\mathbf{p}^{n-1}Q\big|_{boundary} = 0 \tag{B7}$$

So, the quantum Boltzmann equation (23) is integrated with respect to the momentum

$$\frac{\partial}{\partial t}\int f(\mathbf{r},\mathbf{p},t)d^3\mathbf{p} + \frac{1}{m}\nabla.\int f(\mathbf{r},\mathbf{p},t)\mathbf{p}d^3\mathbf{p} + \int \mathbf{F}.\nabla_p f(\mathbf{r},\mathbf{p},t)d^3\mathbf{p}$$
$$= -\frac{\hbar}{2m}\nabla^2 \int \text{Im}Q(\mathbf{r},\mathbf{p},t)d^3\mathbf{p} + \sum_{n=2}^{\infty} O(n) \tag{B8}$$

The sum at the right hand side indicates the contribution of the last two terms of the equation (23). By virtue of relations (B6)-(B7), the first two terms only gives nonvanishing contribution, so

$$\frac{\partial}{\partial t}\rho + \nabla \cdot (\rho \mathbf{u}) = 0 \tag{B9}$$

This is the continuity equation.

**B2. Conservation of momentum**

To obtain momentum conservation relation, the quantum Boltzmann equation is multiplied by momentum vector **p** and then integrated

$$\frac{\partial}{\partial t}\int f(\mathbf{r},\mathbf{p},t)\mathbf{p}d^3\mathbf{p} + \frac{1}{m}\nabla.\int f(\mathbf{r},\mathbf{p},t)\mathbf{pp}d^3\mathbf{p} + \int \mathbf{F}.\nabla_\mathbf{p} f(\mathbf{r},\mathbf{p},t)\mathbf{p}d^3\mathbf{p}$$
$$= -\frac{\hbar}{2m}\nabla^2 \int \text{Im}Q(\mathbf{r},\mathbf{p},t)\mathbf{p}d^3\mathbf{p} + \sum_{n=2}^{\infty} O(n) \tag{B10}$$

for $n \geq 2$

$$O(n) \approx (\text{Re or Im})\int \nabla_{\mathbf{p}}^n Q(\mathbf{r},\mathbf{p},t)\mathbf{p}d^3\mathbf{p} \approx (\text{Re or Im})\int \nabla_{\mathbf{p}}^{n-1} Q(\mathbf{r},\mathbf{p},t)d^3\mathbf{p} = \mathbf{0} \quad (B11)$$

Each term of (B10) is substituted by the help of the above derived relations (B1) – (B7)

$$\frac{\partial}{\partial t}(\rho\hbar\nabla\varphi) + \nabla\cdot[m\rho(\frac{\hbar\nabla\varphi}{m})(\frac{\hbar\nabla\varphi}{m})] - \frac{\hbar^2}{m}\nabla\cdot[\rho^{1/2}\nabla(\frac{1}{2}\rho^{-1/2}\nabla\rho)] - \rho\mathbf{F}$$
$$= -\frac{\hbar^2}{4m}\nabla(\nabla^2\rho) \quad (B12)$$

The last term in (B12) is combined with the third term to give

$$-\frac{\hbar^2}{m}\nabla\cdot[\rho^{1/2}\nabla(\frac{1}{2}\rho^{-1/2}\nabla\rho)] + \frac{\hbar^2}{4m}\nabla(\nabla^2\rho)$$

$$= -\frac{\hbar^2}{2m}\rho\nabla[\frac{1}{2\rho}(\nabla^2\rho - \frac{(\nabla\rho)^2}{2\rho})] = \rho\nabla\varepsilon \quad (B13)$$

Inserting (B13) into (B12) gives the momentum conservation relation

$$\partial_t[m\rho(\frac{\hbar\nabla\phi}{m})] + \nabla\cdot[m\rho(\frac{\hbar\nabla\phi}{m})(\frac{\hbar\nabla\phi}{m})] - \rho\mathbf{F} + \rho\nabla\varepsilon = \mathbf{0} \quad (B14)$$

This is exactly the equation (6).

**B3. Energy conservation**

To derive energy conservation relation, at first some relations are proved

$$\int \frac{p^2}{2m}f(\mathbf{r},\mathbf{p},t)d^3\mathbf{p} = \text{Re}[-\frac{\hbar^2}{2m}\psi(\mathbf{r},t)\nabla^2\psi^*(\mathbf{r},t)] = \rho[\frac{(\hbar\nabla\varphi)^2}{2m} + \varepsilon] \quad (B15)$$

$$\int (p^2/2m)(\mathbf{p}/m)\cdot\nabla f(\mathbf{r},\mathbf{p},t)d^3\mathbf{p} = \nabla\cdot\int (p^2/2m^2)\mathbf{p}f(\mathbf{r},\mathbf{p},t)d^3\mathbf{p}$$
$$= \text{Re}\,\nabla\cdot[(i\hbar)^3\psi\nabla^2\nabla\psi^*]/(2m^2) \quad (B16)$$

$$\int (p^2/2m)\mathbf{F}\cdot\nabla_{\mathbf{p}}fd^3p = -(\mathbf{F}/m)\cdot\int \mathbf{p}fd^3\mathbf{p} = -\rho\mathbf{F}\cdot\mathbf{u} \quad (B17)$$

$$\int (\mathbf{p}^2/2m)\nabla^2(\text{Im}Q)d^3\mathbf{p} = \text{Im}[\nabla^2\int (\mathbf{p}^2/2m)Qd^3\mathbf{p} = \text{Im}\nabla^2[\frac{(i\hbar)^2}{2m}(\psi\nabla^2\psi^*)] \quad (B18)$$

For $n \geq 2$

$$\int \frac{p^2}{2m}\nabla_{\mathbf{p}}^n \text{Im}\,Q(\mathbf{r},\mathbf{p},t)d^3\mathbf{p} = \frac{1}{2m}(\nabla_{\mathbf{p}}\nabla_{\mathbf{p}}p^2)\int \nabla_{\mathbf{p}}^{n-2}Qd^3\mathbf{p} = \mathbf{0} \quad (B19)$$

For $n \geq 3$

$$\int \frac{p^2}{2m}\nabla_{\mathbf{p}}^n f(\mathbf{r},\mathbf{p},t)d^3\mathbf{p} = \frac{1}{2m}(\nabla_{\mathbf{p}}\nabla_{\mathbf{p}}p^2)\int \nabla_{\mathbf{p}}^{n-2}fd^3\mathbf{p} \equiv f|_{boundary} = \mathbf{0} \quad (B20)$$

So, the contribution of two last sums in (23) vanishes. The definition (2) permits to express (B16) and (B18) versus $\rho$ and $\phi$. The final result is

$$\frac{\partial}{\partial t}(\rho\varepsilon + \frac{1}{2}m\rho\mathbf{u}^2) + \nabla\cdot\{\rho\mathbf{u}(\varepsilon + \frac{1}{2}m\rho\mathbf{u}^2) + \frac{\hbar^2}{4m}\{\frac{1}{\rho}\nabla[\rho\nabla\cdot(\rho\mathbf{u})]\}\} - \frac{\mathbf{F}}{m}\cdot\rho\mathbf{u} = 0 \quad (B21)$$

where the average velocity $\mathbf{u}$ is already defined in Eq. (7). The same result is derived by direct differentiation of the first term in (B21) and by the help of the previously obtained conservation relations, where the definition of $\varepsilon$ per $\rho$ should be used. The above equation is quit similar to the classical energy conservation relation (A3).

## APPENDIX C: DERIVING THE BOLTZMANN RELATION FROM LIOVILLE EQUATIONS

A classical system, containing $N$ particles, is described by a point in the $6N$ phase space, which is constructed from $3N$ coordinates for each space and momentum. Suppose that there are $n$ such systems in the elemental volume $\Delta V$. These systems will be contained in some other elemental volume $\Delta V'$, under evolution at some later time. Therefore, the conversation of the systems under time evolution would require

$$D^{(N)}\Delta V = D'^{(N)}\Delta V' \tag{C1}$$

in which, the $D^{(N)}$ represents the systems density in phase space. It is shown that the time evolution does not change the volume

$$\Delta V = \Delta V' \tag{C2}$$

It is simply proved by the help of three assumptions

$$\frac{d\mathbf{r}}{dt} = \mathbf{p}/m, \quad \frac{d\mathbf{p}}{dt} = \mathbf{F}, \quad \frac{\partial}{\partial p} \cdot \mathbf{F} = o \tag{C3}$$

The combined (C1) and (C2) gives the Liouville equation

$$\frac{D}{Dt}D^{(N)} = [\frac{\partial}{\partial t} + \sum_i \frac{\mathbf{p_i}}{m} \cdot \frac{\partial}{\partial r_i} + \sum_i \frac{d\mathbf{p_i}}{dt} \cdot \frac{\partial}{\partial p_i}]D^{(N)} = 0 \tag{C4}$$

where the summation is made over all particles. By defining

$$F_i^{(1)}(x_i,t) = \int D^{(N)} \prod_j dx_j, j \neq i \tag{C5}$$

$$F_{ij}^{(2)}(x_i,x_j,t) = \int D^{(N)} \prod_k dx_k, k \neq i,j \tag{C6}$$

the following equation is obtained

$$(\frac{\partial}{\partial t} + \frac{\mathbf{p_i}}{m} \cdot \frac{\partial}{\partial r_i} + \mathbf{F_i} \cdot \frac{\partial}{\partial p_i})F_i^{(1)}(x_i,t) + \sum_{j \neq i}^N F_{ij} \cdot \frac{\partial}{\partial p_i} F_{ij}^{(2)} dx_j = 0 \tag{C7}$$

$F_i$ is the force exerting on particle $i$ while, $F_{ij}$ is the mutual force of two particles $i$ and $j$. What we know as Boltzmann distribution function $f(x,t)$ is an average, over a space domain, of $F_i^{(1)}(x_i,t)$

$$\frac{1}{m^3}f(x,t)\Delta x = \sum_{i=1}^N \int_x^{x+\Delta x} F_i^{(1)}(x_i,t)dx_i \tag{C8}$$

After a number of assumptions in kinetics, coarse graining etc., equation (C7) yields the Boltzmann equation [11]:

$$(\frac{\partial}{\partial t} + \frac{\mathbf{p}}{m} \cdot \frac{\partial}{\partial \mathbf{r}} + \mathbf{F} \cdot \frac{\partial}{\partial \mathbf{p}})f(x,t) + J = 0 \tag{C9}$$

The collision term $J$ reflects the binary interaction. It is absent for a system of one particle.

---

[1] A. Einstein, B. Podolvski, and N. Rosen, Physical Review **47**, 777 (1935).

[2] J. S. Bell, Reviews of Modern Physics **38**, 447 (1966).

[3] A. Aspect, P. Grangier, and G. Roger, Physical Review Letters **47**, 460 (1981).

[4] E. Schrödinger, Naturwissenschaftern **23**, 807 (1935).

[5] John von Neumann, *Mathematical foundations of quantum mechanics* (Princeton University Press, New York, 1955).

[6] D. Bohm, Physical Review **85**, 166 (1952).

[7] H. Everett, Review of Modern Physics **29**, 454 (1957).

[8] G. C. Ghirardi, A. Rimini, and T. Weber, Physical Review D **34**, 470 (1986).

[9] J. G. Cramer, Physical Review D **22**, 362 (1980).

[10] J. Baggott, *The meaning of quantum theory* (Oxford University Press, Oxford, 1992).

[11] T. Koga, *Introduction to kinetic theory stochastic processes in gaseous systems* (Pergamon Press, Oxford, 1970).